\documentclass[10pt]{article}
\usepackage[OE]{express}
\usepackage{color}

\begin{document}

\title{High-quality trapped modes in all-dielectric metamaterials}

\author{Vladimir~R.~Tuz,\authormark{1,2,3,*} Vyacheslav~V.~Khardikov,\authormark{1,3,5} Anton~S.~Kupriianov, \authormark{1,4} Kateryna~L.~Domina,\authormark{5} Su~Xu,\authormark{2} Hai~Wang,\authormark{2} and Hong-Bo~Sun\authormark{2}}

\address{\authormark{1}International Center of Future Science, Jilin University, 2699 Qianjin Street, Changchun 130012, China\\
\authormark{2}State Key Laboratory on Integrated Optoelectronics, College of Electronic Science and Engineering, Jilin University, 2699 Qianjin Street, Changchun 130012, China\\
\authormark{3}Institute of Radio Astronomy of
National Academy of Sciences of Ukraine, 4, Mystetstv Street, Kharkiv 61002, Ukraine\\
\authormark{4}College of Physics, Jilin University, 2699 Qianjin Street, Changchun 130012, China\\
\authormark{5}School of Radio Physics, V. N. Karazin Kharkiv National University, 4, Svobody Square, Kharkiv 61022, Ukraine}

\email{\authormark{*}tvr@rian.kharkov.ua} %% email address is required

% \homepage{http:...} %% author's URL, if desired

%%%%%%%%%%%%%%%%%%% abstract and OCIS codes %%%%%%%%%%%%%%%%
%% [use \begin{abstract*}...\end{abstract*} if exempt from copyright]

\begin{abstract}
A planar all-dielectric metamaterial made of a double-periodic lattice whose unit cell consists of a single subwavelength dielectric particle having the form of a disk possessing a penetrating hole is considered. The resonant states in the transmitted spectra of the metamaterial are identified considering modes inherent to the individual cylindrical dielectric resonator. A correlation between the asymmetry in particle's geometry, which arises from the off-centered displacement of the hole, and formation of the Mie-type and trapped modes is established. The advantages of using a coaxial-sector notch instead of a round hole for the trapped mode excitation are explained.
\end{abstract}

\ocis{(160.3918) Metamaterials; (260.5740) Resonance; (290.4020) Mie theory; (310.6628) Subwavelength structures, nanostructures.} % REPLACE WITH CORRECT OCIS CODES FOR YOUR ARTICLE, MINIMUM OF TWO; Avoid using the OCIS codes for “General” or “General science” whenever possible.
%For a complete list of OCIS codes, visit: https://www.osapublishing.org/oe/submit/ocis/

%%%%%%%%%%%%%%%%%%%%%%% References %%%%%%%%%%%%%%%%%%%%%%%%%

%%%%%%%%%%%%%%%%%%%%%%%%%%  body  %%%%%%%%%%%%%%%%%%%%%%%%%%
\section{Introduction}
As an alternative method for designing electromagnetic metamaterials, the dielectric particles show several advantages, notably in terms of isotropic parameters, low losses, and the fabrication techniques targeting at higher frequencies of operation \cite{Zhao_MatToday_2009, jahani_NatNano_2016, Kivshar_NatPhot_2017}. In such all-dielectric metamaterials subwavelength dielectric particles made of high-$\varepsilon$ materials are arranged into a lattice where each particle behaves as an individual resonator sustaining a set of electric and magnetic multipole modes whose coupling to the field of the incident wave produces a strongly resonant electromagnetic response of the metamaterial. Desirable overlapping of certain multipole resonances (which are usually referred to `Mie-type' modes) produced by the dielectric particles can be used for different applications, including nanoantennas \cite{Krasnok_OptExpress_2012, Baryshnikova_JOptSocAmB_2017}, sensors \cite{Bontempi_Nanoscale_2017}, solar cell technology \cite{Spinelli_NatComm_2012}, and multifunctional metasurfaces \cite{Decker_AdvOptMat_2015}. 

Among a variety of known metamaterial configurations we further distinguish a particular class of planar resonant structures which allow obtaining the strongest resonant response due to the excitation of so-called `trapped modes' \cite{Zouhdi_Advances_2003, Fedotov_PhysRevLett_2007} (in literature this type of resonant excitations is also referred to `dark states' \cite{Zhang_PhysRevLett_2008, Jain_AdvOptMater_2015}). Such modes appear in metamaterials provided that their subwavelength particles possess certain structural asymmetry. The degree of this asymmetry determines the strength of electromagnetic coupling between the incident field and currents induced by this field inside the metamaterial particles. In order to achieve the trapped mode excitation in all-dielectric metamaterials a particular \textit{two-particle} design is proposed \cite{Khardikov_JOpt_2012, Khardikov_NATO, Zhang_OptExpress_2013, Zhang_OptExpress_2014, Prosvirnin_ApplOpt_2015} where two dissimilar dielectric bodies (parallelepipeds) made of a low-loss high-$\varepsilon$ material are combined together to form a periodic unit cell. In such a design each particle serves as an individual dielectric resonator whereas an electromagnetic coupling between them caused by antiphased \textit{displacement} (polarization) currents induces the field concentration within the system. Since at the trapped mode resonant frequency the electromagnetic field is strongly confined inside the system there prospects appear to design highly desirable tunable deep-sub-wavelength planar metamaterials that can provide a more efficient employment of materials which exhibit pronounced absorption, nonlinear characteristics or properties of gain media \cite{Zheludev_NatPhoton_2008, tuz_PhysRevB_2010, tuz_EurPhysJApplPhys_2011, tuz_JOpt_2012, Khardikov_RadioPhysAstron_2013, Khardikov_ACSPhot_2015, Khardikov2016, Liu_OptExpress_2017}.

In the present paper it is our goal to reveal and classify the trapped modes whose resonant conditions are achieved in a \textit{single-particle} all-dielectric planar metamaterial. Thus, we demonstrate a very simple design of an all-dielectric metamaterial whose lattice consists of only a single dielectric body per unit cell. Each particle within the lattice behaves as an individual dielectric resonator supporting a set of electric and magnetic modes. We show that if a particular asymmetry is introduced inside the unit cell of the structure, in addition to common electric and magnetic Mie-type resonances, the trapped mode with an unprecedented quality factors can be excited. Considering the characteristics of trapped mode excitation we propose a special design of the unit cell where a short coaxial-sector notch (smile) is made through the disk that has a minimal impact on the conditions of excitation of the dipole Mie-type modes compared to those existed in the commensurable solid disk.  

\section{\label{sec:design}Design and simulation}

\begin{figure}[ht!]
\centering\includegraphics[width=7cm]{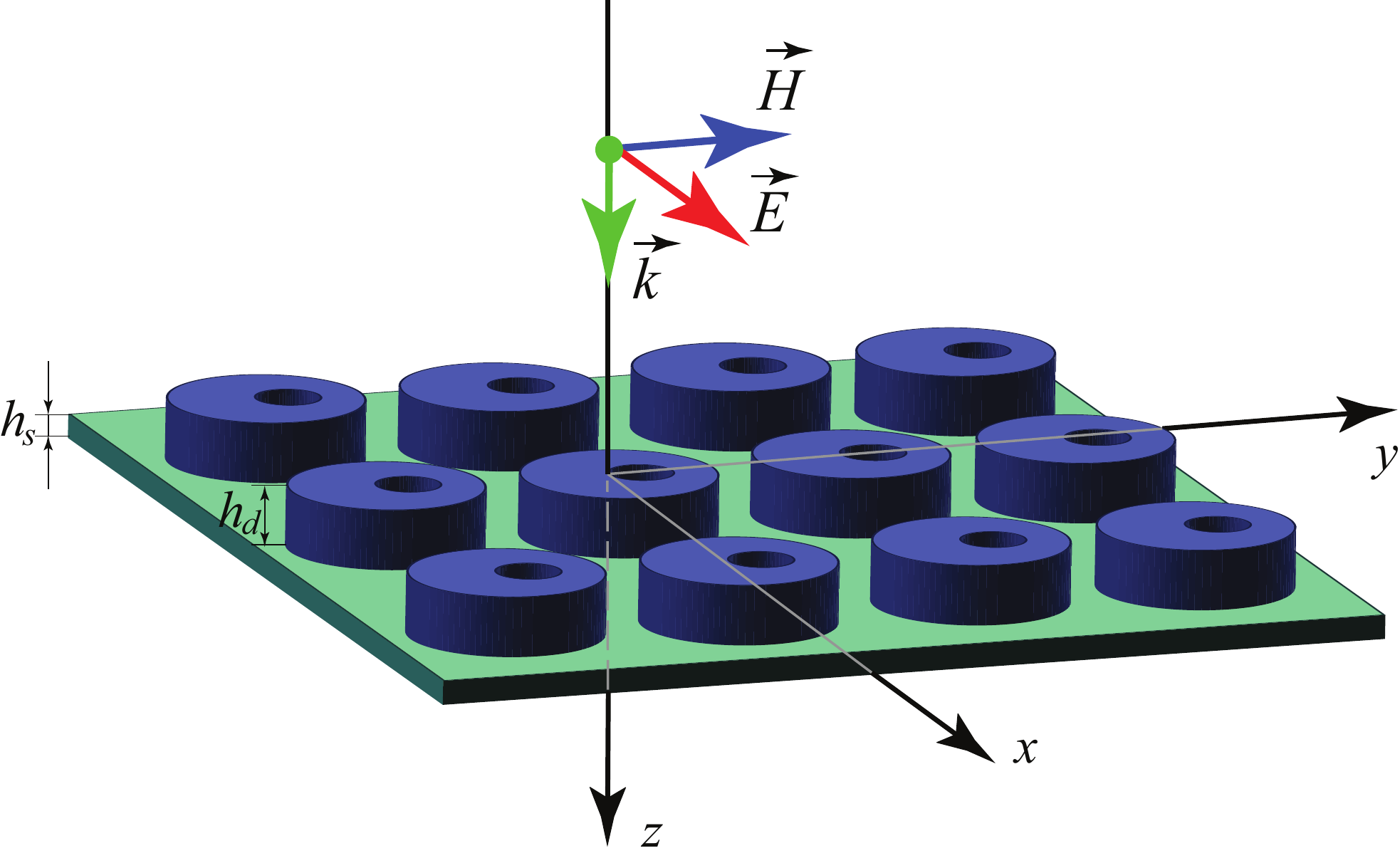}
\caption{Fragment of a single-particle planar all-dielectric metamaterial. The structure is under an illumination of a normally incident plane wave whose electric field vector is directed along the $x$-axis ($x$-polarization).}
\label{fig:struct}
\end{figure}

Thereby, further we carry out a comparative study of resonant characteristics of planar all-dielectric metamaterials which differ by a construction of the unit cell. In particular, we study double-periodic in the $x-y$ plane lattice with a square unit cell ($d_x = d_y = d$) in which a single circular disk with radius $a_d$ and height $h_d$ is deposited (Fig.~\ref{fig:struct}). In addition to a solid disk, we consider those that have either a centered or off-centered penetrating hole (notch). In the latter case the particle's design is chosen in such a way that the unit cell appears to be symmetric with respect to the line drawn through its center parallel to the $y$-axis, whereas with respect to the $x$-axis it is asymmetric. The disks are made from a nonmagnetic lossless dielectric having permittivity $\varepsilon_d$. Finally, the lattice is imposed on top of a thin dielectric substrate (thin compared to the wavelength of incident field) having permittivity $\varepsilon_s$ and thickness $h_s$ to form a whole metamaterial.

It is our main goal to excite a particular resonant state known as the trapped mode which arises when particles forming a metamaterial possess certain structural asymmetry \cite{Zouhdi_Advances_2003, Fedotov_PhysRevLett_2007}. In the planar structure under study such an asymmetry appears in the $x-y$ plane with respect to the $x$-axis, and for the trapped mode to be excited, the vector of electric field strength $\vec E$ must have the component directed along this axis. Therefore, further we define that the metamaterial is illuminated by a normally incident plane wave $(\vec k = \{0,0,k_z\})$ whose electric field vector $(\vec E=\{E_x,0,0\})$ is directed along the $x$-axis ($x$-polarization).

Unlike planar metamaterials comprising a lattice of metallic particles, where the resonant frequency is defined by the total length of particle arcs because the \textit{surface} currents flow along these arcs following their complicate shape \cite{Fedotov_PhysRevLett_2007}, in an all-dielectric metamaterial each single particle behaves as an open volumetric dielectric resonator where the \textit{displacement} currents excited by the incident wave form the corresponding inner electromagnetic field pattern. In order to reduce the operating frequency of such a resonator one needs either to increase the volume of dielectric body, or use some high-contrast optical materials for the resonator fabrication. Moreover, according to the main principle of metamaterial theory, the particle transverse dimensions are restricted by the size of unit cell which must be less than the wavelength of the incident field (i.e. it should be \textit{subwavelength}). Thus, the high--$\varepsilon$ dielectrics appear to be the most appropriate materials for the fabrication of particles of the double-periodic lattice in the all-dielectric metamaterial under study.

Among such materials certain semiconductors are considered to be suitable for fabrication resonant particles in the infrared and visible parts of spectrum. Within these wavelengths there are corresponding transparency windows where semiconductors acts as good high-$\varepsilon$ dielectrics having small dielectric loss tangent which does not usually exceed $10^{-3}$. In particular, a lot of metamaterial designs based on arrays of silicon, gallium arsenide or germanium particles have been already proposed \cite{Zhang_OptExpress_2013, Grzegorczyk_PhysRevLett_2007, Cummer_PhysRevLett_2008, Zywietz_NatCommun_2014}.

Besides, in order to manufacture all-dielectric metamaterials operating in the microwave range an inexpensive bottom-up-chemical approach utilizing commercially available ceramic materials can be used. Thus, it is proposed to use a fine powder of titanium dioxide \cite{Yahiaoui_ApplPhysLett_2012} for the particles fabrication, since titanium dioxide excellently matches requirements having very high permittivity and moderate low loss tangent \cite{Wypych_JNanomat_2014}. According to such a technology dielectric resonators are produced by pressing the powder of pure titanium dioxide into pellets and then pellets sintering in a furnace with subsequent assembling obtained dielectric particles into a periodic lattice. The matrix for lattice holding can be made by mechanically drilling holes in a dielectric host substrate or be deposited by 3D printing technique. In order to  satisfy the resonant conditions, a high refractive index contrast between the disks and matrix must be ensured. Therefore, further in this study we consider a metamaterial prototype whose lattice is made from titanium dioxide ($\varepsilon_d=57$) disks arranged into a 3D printed ABS matrix ($\varepsilon_s=1.2$) for operating in the $40-70$~GHz band. 

Our numerical simulations of the electromagnetic field scattering by a double-periodic lattice of volumetric dielectric resonators are based on both original and commercial software. Developed original approaches and codes exploit the pseudo-spectral method in time domain \cite{Khardikov_RadioPhysAstron_2008}, whereas validation of the original codes and particular calculations of the resonant electric and magnetic field distribution inside the dielectric particles were performed using the commercial COMSOL Multiphysics finite-element-based electromagnetic solver.

\section{\label{sec:results}Transmitted spectra and mode classifications}

In order to reveal main peculiarities of the trapped mode excitation, in this section we consistently consider an electromagnetic response (transmitted spectra) of three particular designs of a planar all-dielectric metamaterial whose unit cell consists of either symmetric or asymmetric inclusions. Regardless of the symmetry property, the resonant conditions arising in the metamaterial under study primarily depend on the modal composition of individual dielectric resonators forming the metamaterial rather than on the electromagnetic coupling of these resonators within the whole lattice (see Appendix A). Therefore, considering the inner electromagnetic field pattern and displacement currents distribution inside the unit cell at particular resonant frequencies we uniquely associate the obtained resonant states with corresponding electric or magnetic modes which are inherent to individual cylindrical resonators forming the metamaterial.

\subsection{\label{sec:symmetricsolid}Symmetric disk: Mie-type modes}

\begin{figure}[ht!]
\centering\includegraphics[width=10cm]{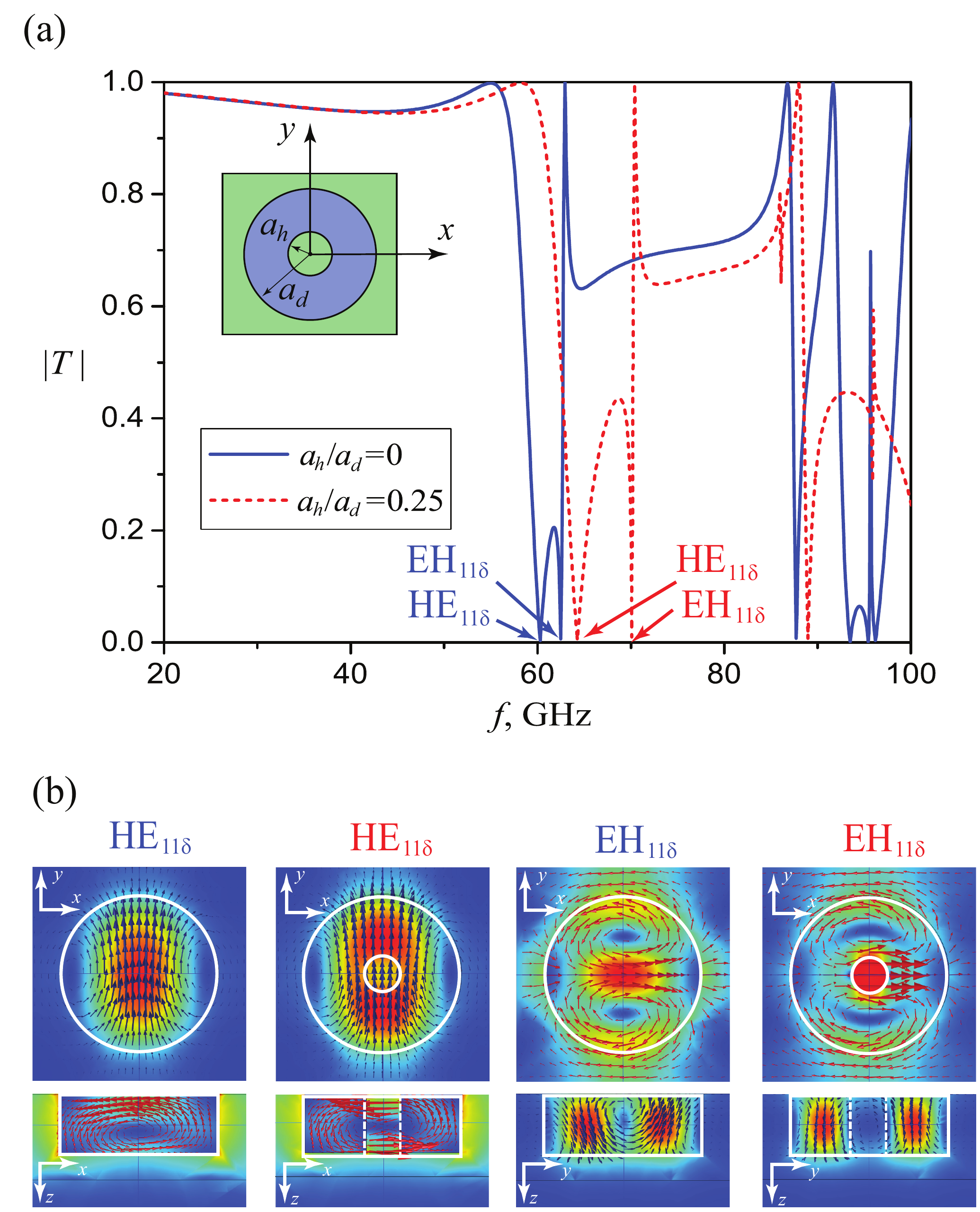}
\caption{(a) Transmission coefficient magnitude of an all-dielectric metamaterial possessing symmetrical unit cell with a solid disk (solid blue line) and a disk having centered ($\theta=0$) round hole (dashed red line), and (b) cross-section patterns of electric (red arrows) and magnetic (blue arrows) field distribution which are calculated within the unit cell at the corresponding resonant frequencies of Mie-type modes; $a_d=0.457$~mm, $h_d=0.417$~mm, $d=1.25$~mm, and $h_s=0.167$~mm.}
\label{fig:symmetric}
\end{figure}

In order to ensure the completeness of our study, we start the discussion by demonstrating an electromagnetic response of the metamaterial whose unit cell possesses a \textit{symmetric} design (see also Refs.~\cite{Decker_AdvOptMat_2015, Liu_OptExpress_2017}). Thereby, the calculated transmitted spectra of the metamaterial comprising of solid disks without holes and disks having centered round penetrating hole (with radius $a_h$) are plotted in Fig.~\ref{fig:symmetric}(a) with solid blue and dashed red lines, respectively. In the frequency band of interest two resonant states are distinguished and marked at the bottom of the figure with blue and red arrows. These resonant states acquire a sharp peak-and-trough (resonance-antiresonance) profile \cite{Kivshar_NatPhot_2017}. While the dip in curves corresponds to the maximum of reflection, the peak corresponds to the maximum of transmission. These extremes approach to 0 and 1, respectively, since the losses in the given structure are considered to be negligibly small.

From the analysis of the cross-section patterns of the electric and magnetic field and displacement current distributions presented in Fig.~\ref{fig:symmetric}(b) one can conclude that the identified resonances appear due to an existing electromagnetic coupling between the linearly polarized incident wave and the lowest-order (dipole) magnetic (HE$_{11\delta}$; first Mie resonance \cite{Bohren_2008}) and electric (EH$_{11\delta}$; second Mie resonance) hybrid modes inherent to the individual cylindrical resonator (on the modes existing in cylindrical waveguides and resonators see, for instance, Refs.~\cite{Snitzer_JOptAm_61, marcatili1964hollow, mongia_IntJRFMicrowave_1994}). In the mode denotation the subscripts correspondingly define alternating numbers of the electromagnetic field variations inside the resonator along the azimuthal, radial and longitudinal directions. In the third subscript the index $p+\delta$ ($p=0,1,...$) is also introduced to define the number of half-wavelengths standing along the $z$-axis, where $\delta=(2h_d/\lambda_g - p)<1$, and $\lambda_g$ is the wavelength of corresponding mode of the cylindrical dielectric resonator \cite{mongia_IntJRFMicrowave_1994}.

For the given polarization of the incident wave, the resonant pattern of the electric field establishes in the longitudinal ($x-z$) or ($y-z$) plane and transverse ($x-y$) plane for the HE$_{11\delta}$ mode and EH$_{11\delta}$ mode, respectively. Therefore, it is easy to conclude that, among other resonator parameters, the resonant frequency of the HE$_{11\delta}$ mode is mainly determined by the disk height, whereas that of the EH$_{11\delta}$ mode depends primarily on the disk radius. That is why the order of these resonances on the frequency scale may be different depending on the height and radius of the resonator. Apart from the lowest-order (dipole) modes some higher-order (multipole) Mie-type modes are also excited in the given metamaterial. However, since they are lying in the higher frequency band which is out of interest they are excluded from our further consideration.

If a cylindrical dielectric plug is removed from near the axis of the disk, the resulting cylindrical ring resonator remains supporting the same type of modes as a solid one \cite{mongia_IntJRFMicrowave_1994}. All previously identified resonant states arise also in the transmitted spectra of the modified structure. However, since the resonant frequencies of both modes of the individual ring resonator becomes greater than those of the corresponding commensurable solid disk, the resonant states in the transmitted spectra of the modified metamaterial acquire some frequency shift. This shift occurs due to the fact that the created hole generally reduces the effective permittivity of the dielectric resonator. It results also in the stronger frequency shift of the resonant state related to the EH$_{11\delta}$ mode considering the hole is located directly in the maximum of the electric field in this case.

\subsection{\label{sec:asymmetrichole}Asymmetric hole: Trapped modes}

\begin{figure}
\centering\includegraphics[width=10cm]{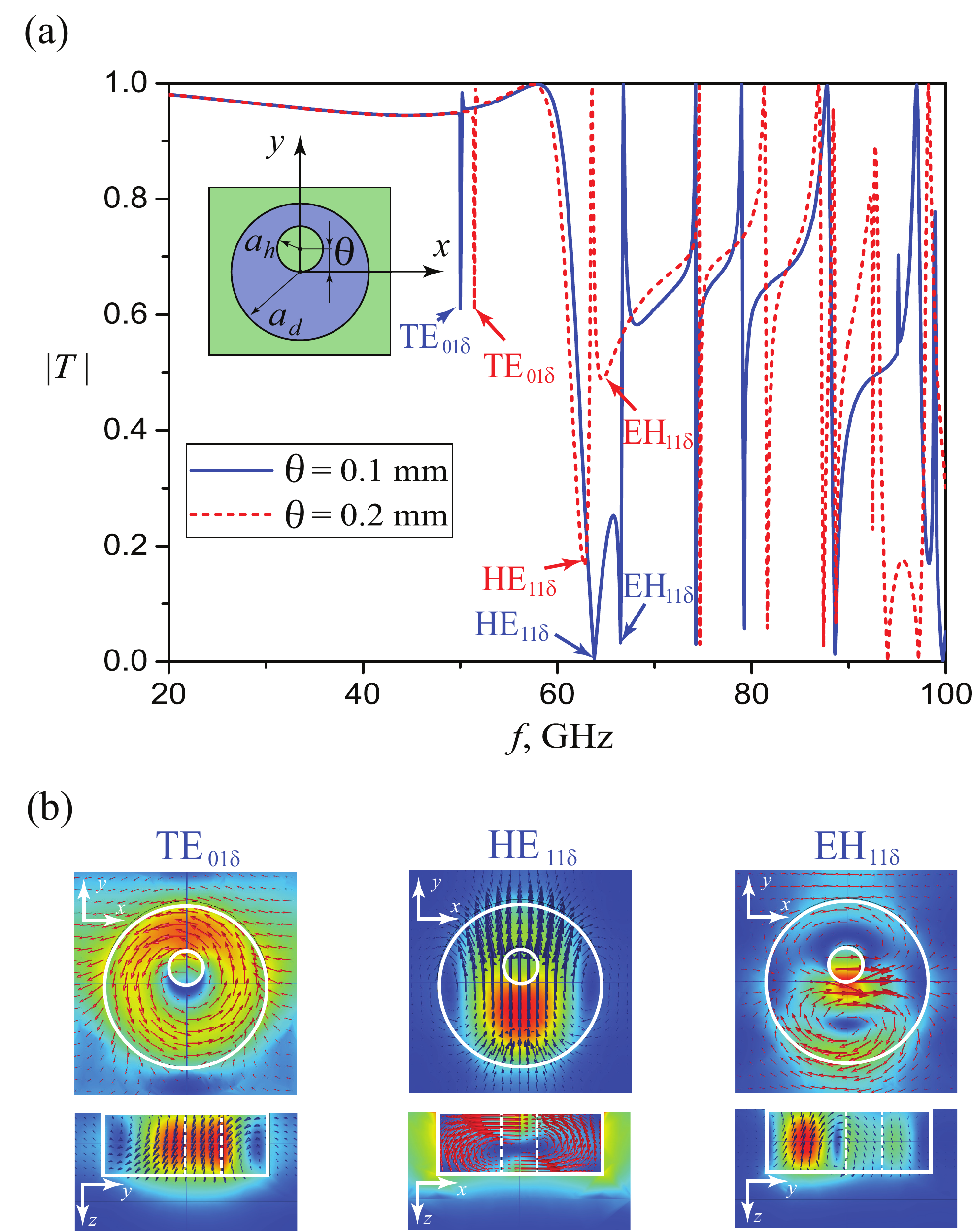}
\caption{(a) Transmission coefficient magnitude of an all-dielectric metamaterial possessing asymmetrical unit cell with a disk having off-centered ($\theta\ne 0$) round hole, and (b) cross-section patterns of electric (red arrows) and magnetic (blue arrows) field distribution which are calculated within the unit cell at the corresponding resonant frequencies of Mie-type and trapped modes; $a_d=0.457$~mm, $h_d=0.417$~mm, $d=1.25$~mm, $h_s=0.167$~mm, and $a_h/a_d=0.25$.}
\label{fig:asymmetric}
\end{figure}

In the second metamaterial design each unit cell consists of a disk having a hole shifted along the $y$-axis regarding the disk center on the distance $\theta$ (see also Ref. \cite{Jain_AdvOptMater_2015}). Since in this case an asymmetry in the unit cell is introduced, additional resonant states appear in the transmitted spectra of the metamaterial. For such an \textit{asymmetric} metamaterial two curves of the transmission coefficient magnitude are plotted in Fig.~\ref{fig:asymmetric}(a) which correspond to different values of the lateral displacement of the hole across the disk within the unit cell, assuming the hole radius is fixed. As previously, in this figure all resonant states of interest are distinguished and marked with blue and red arrows. Then the origin of these states is recognized from the electric and magnetic field patterns and displacement current distributions calculated at corresponding resonant frequencies. They are depicted in Fig.~\ref{fig:asymmetric}(b).

One can see that in the asymmetric metamaterial all previously discussed resonant states related to the lowest-order Mie-type modes generally retain their positions on the frequency scale compared to the spectra of the symmetric metamaterial. As the hole displacement arises, the resonant state related to the EH$_{11\delta}$ mode acquires a shift towards the lower frequencies because the hole leaves the region where the electric field has its maximum. For the same reason, the presence of a hole and its asymmetric displacement have little effect on the position of the HE$_{11\delta}$ mode. Besides, the transmission coefficient magnitude in the frequency band of both Mie-type modes changes in such a way that the reflectivity of the structure decreases. 

\begin{figure}
\centering\includegraphics[width=8cm]{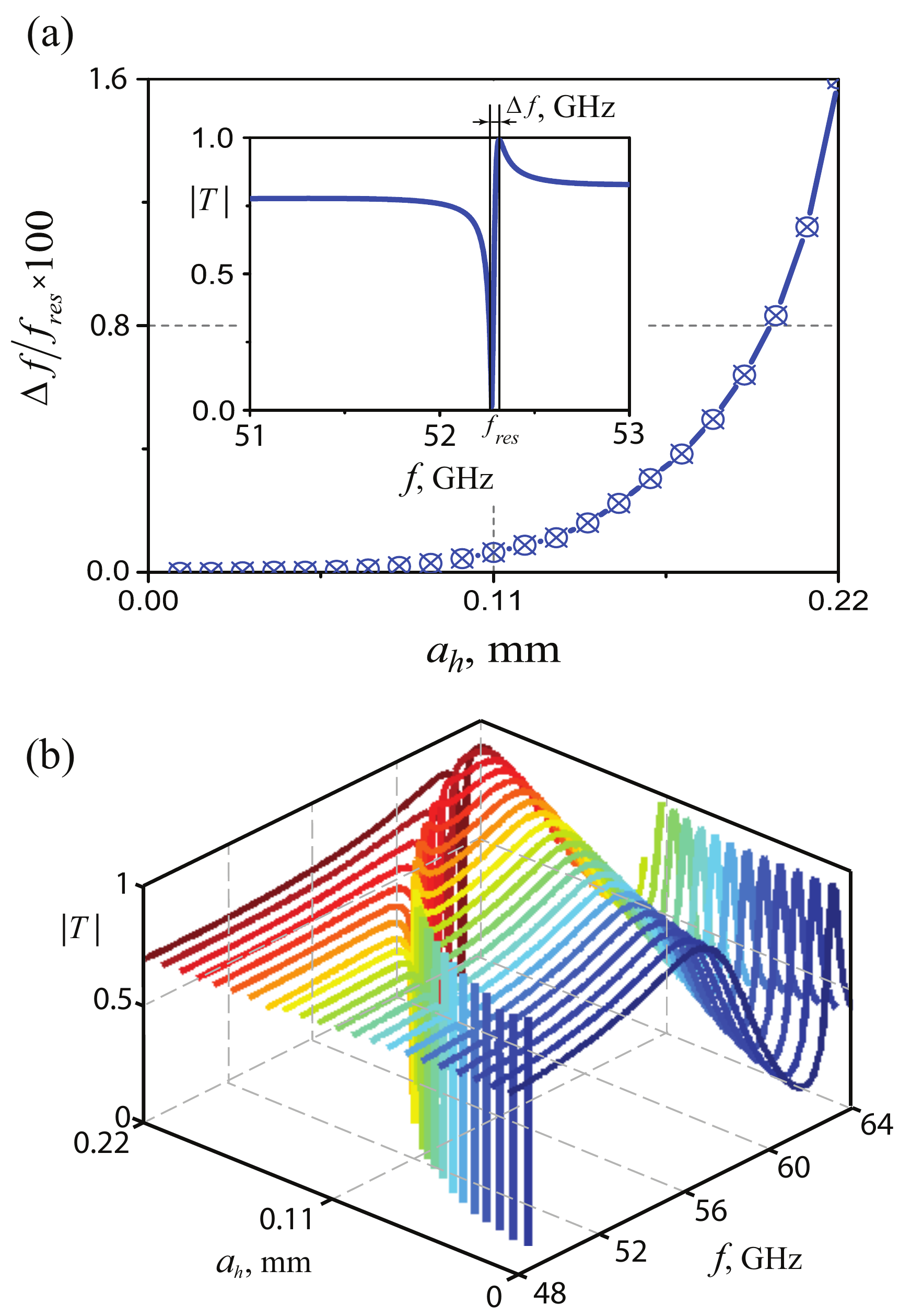}
\caption{(a) Inverse value of the quality factor of the resonant state related to the trapped (TE$_{01\delta}$) mode versus the hole radius when the hole disposition is fixed  and (b) transmission coefficient magnitude as a function of frequency and hole radius; $a_d=0.457$~mm, $h_d=0.417$~mm, $d=1.25$~mm, $h_s=0.167$~mm, and $\theta=0.2$.}
\label{fig:quality}
\end{figure}

Aside from the resonant states related to the above considered Mie-type modes which possess a good electromagnetic coupling with the field of the normally incident linearly polarized wave, in the frequency band of interest an additional resonance arises that is inherent only for the asymmetric design. From the cross sectional pattern of the inner electromagnetic field calculated for this resonance one can conclude that the arrows of the displacement currents demonstrate a circular behavior twisting around the center of the unit cell in the \textit{plane} of the metamaterial which produces a magnetic moment oriented along the $z$-axis. Such a form of the electric field distribution is typical for the transverse electric (TE) modes inherent to the individual cylindrical dielectric resonator \cite{Snitzer_JOptAm_61, marcatili1964hollow, mongia_IntJRFMicrowave_1994}. In particular, considering the number of longitudinal field variations inside the disk this resonant state is related to the TE$_{01\delta}$ mode of the resonator. It is important to note that this mode appears as an alone-standing lowest frequency resonance, which gives an advantage of its utilization, for instance, for sensory applications. 

The presence of displacement currents flowing in a round way leads to the appearance of antiphased components parallel to the $x$-axis which have a very low level of electromagnetic coupling with the field of the linearly $x$-polarized incident wave. In other words, the scattered electromagnetic far-field produced by such displacement current distributions is very weak, which drastically reduces the resonators coupling to free space and therefore reduces strongly the radiation losses. Such a resonant regime is referred to trapped mode, since this term is traditionally used in describing electromagnetic modes which are weakly coupled to free space \cite{Stupakov_PhysRevEB_1994, Kirilenko_MTT_2000, Annino_PhysRevB_2006}. The strength of the induced inner field in the trapped mode can reach very high values which ensure a strong resonant response. Similar to characteristics of the dipole Mie-type modes the spectral line of the trapped mode acquires a sharp peak-and-trough profile, whereas the quality factor of such a resonance appears to be much higher than that of the dipole Mie-type modes.

However, the notion on the quality factor of the discussed resonant state in the lossless structure requires some special explanation. For the state having resonance-antiresonance profile, the quality factor can be defined as a ratio of the resonant frequency $f_{res}$ and a difference $\Delta f$ between the frequency positions related to the maximum (peak) and minimum (dip) of the transmission coefficient magnitude within the resonance. Since the quality factor of the trapped mode resonance tends to infinity in the case of weakly asymmetric disk, then it is suitable to use the value $\Delta f / f_{res}$ to illustrate $Q$-factor dependence on the asymmetry degree. With such a definition it follows that the greater is the difference between positions of these two frequencies, the lower is the quality factor of the resonance. In particular, this difference is presented in the inset of Fig.~\ref{fig:quality}(a), where the essence of $\Delta f$ is clarified.

In order to study the dependence of the quality factor of the trapped mode resonance on the hole radius additional calculations of the transmitted spectra of the asymmetric metamaterial were performed in the frequency band around the excitation of the TE$_{01\delta}$ mode. They are summarized in Fig.~\ref{fig:quality}(b) where a set of curves of the transmission coefficient magnitude is plotted versus frequency and hole radius considering both the disk radius and hole displacement are fixed. It is evident that the quality factor of this resonance depends crucially on the hole radius since this parameter defines the degree of the introduced asymmetry and, thus, the level of the metamaterial coupling to free space. Therefore, there is an option to adjust both spectral position and quality factor of the trapped mode by geometrical tuning the position and radius of hole within the unit cell while keeping the remaining dimensions of the disk unchanged.

\subsection{\label{sec:smilenotch}Short coaxial-sector notch (smile): Mie-type and trapped modes ensemble}

Presence of an off-centered round penetrating hole in the disk creates a required unit cell's asymmetry which results in the trapped mode excitation in the metamaterial. Unfortunately, in such a design the degree of obtained asymmetry is restricted by an acceptable value of the hole radius that cannot be higher than $a_d$. Another drawback of this design is that if the hole radius becomes larger than $\theta$ ($a_h\geq\theta$) the hole appears in the position where there is a maximum of the electric field of the EH$_{11\delta}$ mode, and thus the hole presence strongly influences this mode appearance. Therefore, in order to overcome these drawbacks the shape of the notch inside the disk should be chosen in such a way as to fit the direction of the lines of the displacement currents exciting a particular mode. Since in the case of excitation of the TE$_{01\delta}$ mode the currents flow in a round way the optimal design should be sought by selecting a suitable coaxial-sector notch. Remarkable, such a form of the hole allows one to achieve maximal asymmetry in the unit cell without introducing any perturbation into the central part of the disk which is important for keeping the resonant conditions of the lowest-order Mie-type modes unchanged. 

\begin{figure}
\centering\includegraphics[width=10cm]{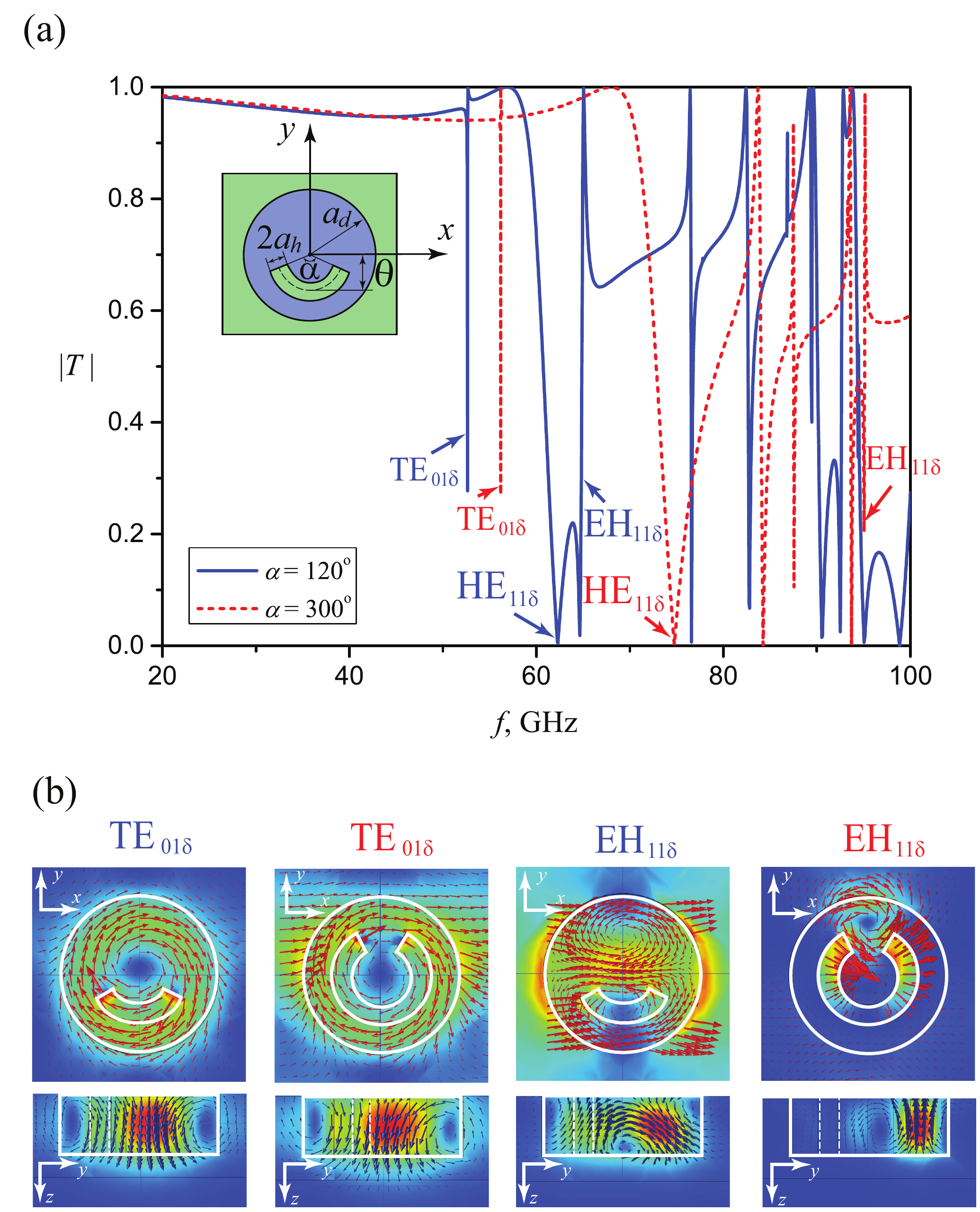}
\caption{(a) Transmission coefficient magnitude of an all-dielectric metamaterial possessing asymmetrical unit cell with a coaxial-sector notch, and (b) cross-section patterns of electric (red arrows) and magnetic (blue arrows) field distribution which are calculated within the unit cell at the corresponding resonant frequencies of TE$_{01\delta}$ and EH$_{11\delta}$ modes; $a_d=0.457$~mm, $h_d=0.43$~mm, $d=1.25$~mm, $h_s=0.167$~mm, $\theta=a_{d}/2$, and $a_{h}/a_{d}=0.125$.}
\label{fig:sector}
\end{figure}

Thus, further in the development of this idea we consider an all-dielectric metamaterial with a lattice whose unit cell consists of a disk possessing a coaxial-sector notch made through it. In this geometry we denote that $\theta$ is the radius of the sector mid-line, $2a_h$ is the notch width, and $\alpha$ is the sector opening angle. The form of the unit cell is presented in the inset of Fig.~\ref{fig:sector}(a) from which one can conclude that in this design the unit cell's asymmetry is preserved with respect to the $x$-axis, while the degree of asymmetry is defined by the sector opening angle $\alpha$. Moreover, it is obvious that the degree of asymmetry varies between two extreme angles $\alpha_{min}=0^\circ$ and $\alpha_{max}=360^\circ$ at which the unit cell becomes symmetric.
 
In order to demonstrate benefits of such a design clearly, the transmitted spectra of the metamaterial whose lattice is made of disks having a coaxial-sector notch are presented in Fig.~\ref{fig:sector}(a) for two different values of the sector opening angle $\alpha$, while the corresponding electric and magnetic field patterns and displacement current distributions calculated at the particular resonant frequencies are depicted in Fig.~\ref{fig:sector}(b). For these two simulations the sector opening angle is selected to be $\alpha=120^\circ$ (short coaxial-sector notch; `smile' disk) and $\alpha=300^\circ$ (long coaxial-sector notch; `spline' disk).  

One can see that for both designs of the unit cell all previously discussed resonant states related to the lowest-order Mie-type and trapped modes appear in the transmitted spectra, and, thus, smile disks as well as spline disks are suitable for the these modes excitation. Nevertheless, while the quality factor and resonant frequency of the trapped modes excited by such disks are almost identical, those of the Mie-type modes are different drastically for these two designs. In fact, since the short coaxial-sector notch appears in the area of local minimum of the electric field of the EH$_{11\delta}$ mode it produces extremely low distortion on the mode appearance. It is not so for the long coaxial-sector notch, because such a sector intersects the $x$-axis along which the lines of the displacement currents of the EH$_{11\delta}$ mode are directed, which reduces effective permittivity of the disk resulting in a stronger high-frequency shift of the mode. For the same reason the resonant frequency of the TE$_{01\delta}$ mode demonstrates the similar behavior. 

\begin{figure}[ht!]
\centering\includegraphics[width=11cm]{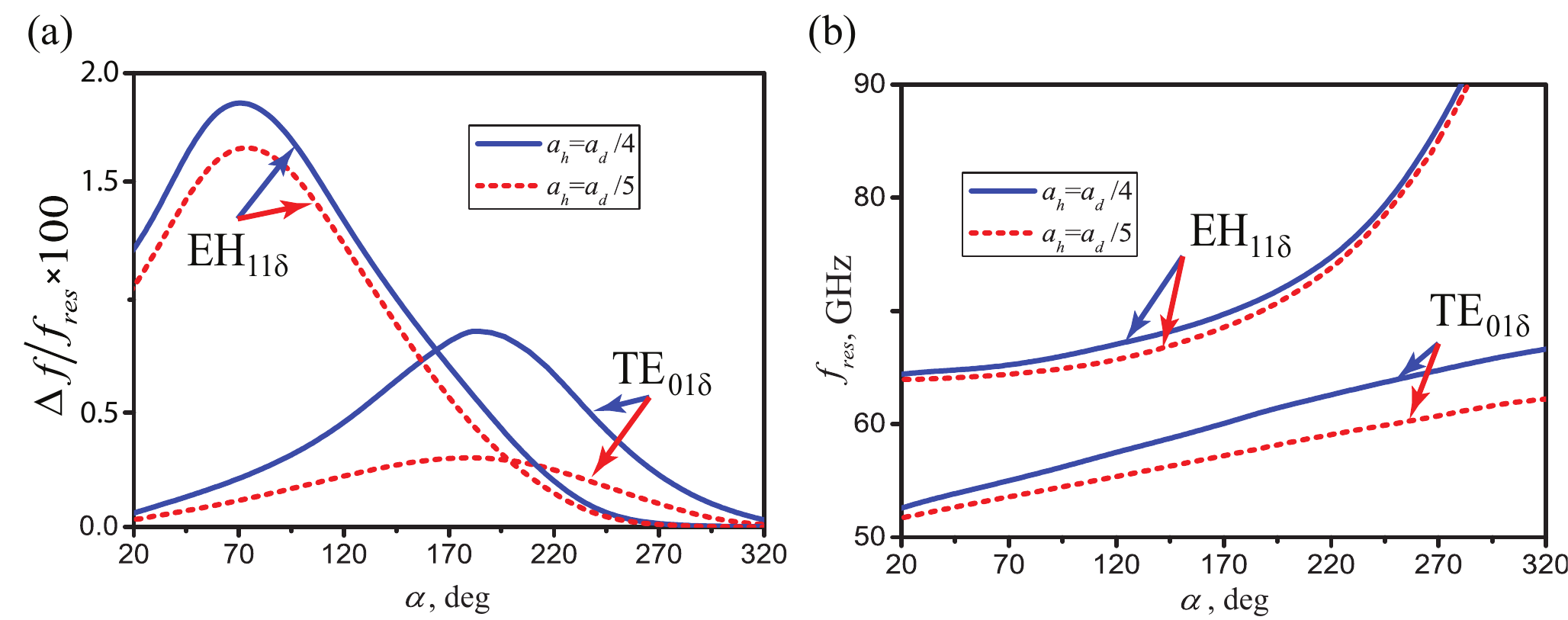}
\caption{(a) Inverse value of the quality factor and (b) resonant frequencies of states related to TE$_{01\delta}$ and EH$_{11\delta}$ modes versus the sector opening angle for two particular values of the notch width; $a_d=0.457$~mm, $h_d=0.417$~mm, $d=1.25$~mm, $h_s=0.167$~mm, and $\theta=a_{d}/2$.}
\label{fig:Qfactor}
\end{figure}

The quality-factor of resonances inevitably depends on the geometric parameters of the notch since they define the degree of the unit cell's asymmetry. Thus, dependence of the quality-factor on the angle $\alpha$ is demonstrated in Fig.~\ref{fig:Qfactor} for both TE$_{01\delta}$ and EH$_{11\delta}$ modes considering two particular values of the notch width $a_h$ while keeping the rest disk parameters fixed. From this figure one can conclude that the quality-factor of the Mie-type mode changes a little with the variation of the notch width, whereas, that of the trapped mode is very sensitive to both notch parameters. Remembering that the resonant frequencies of the lowest-order Mie-type modes are determined by the disk radius and height, tuning the notch parameters opens a prospect to find an optimal solution for the unit cell geometry where the resonant frequencies of the Mie-type modes and trapped modes are brought together as close as possible. In particular, it can be important for the construction of all-dielectric metasurface absorbers \cite{Liu_OptExpress_2017} in view of broadening their operation band. 

An attempt to bring closer the resonant frequencies corresponding to excitations of the trapped mode and electric dipole Mie-type mode of the metamaterial is presented in Fig.~\ref{fig:Qfactor}(b). We should note that the main restriction on these resonances convergence is imposed by the fundamental characteristic of the modes of the individual disk resonator. Thus, the distance between the eigen-frequencies of the TE$_{01\delta}$ mode and EH$_{11\delta}$ mode of the resonator depends only on the disk permittivity and radius, and it cannot be changed by varying other resonator parameters as it is possible, for example, in the case of the EH$_{11\delta}$ mode and HE$_{11\delta}$ mode where one can bring them closer by varying the disk height \cite{Decker_AdvOptMat_2015}. Nevertheless, a coaxial-sector notch of the smile disk reduces the effective permittivity of the resonator for the TE$_{01\delta}$ mode stronger than that for the EH$_{11\delta}$ mode which allows to obtain a certain convergence of their resonant frequencies. From our estimation a distance between these resonant frequencies can be reduced up to 30 percent using disks possessing coaxial-sector notch compared with frequency positions of the corresponding resonances produced by disks with an extremely small off-centered round hole ($a_h=a_{d}/32$).

\section{\label{sec:conclusion}Concluding remarks}
To conclude, despite a very simple design of dielectric inclusions of the proposed all-dielectric metamaterial, in its transmitted spectra along with resonances related to the lowest-order electric and magnetic Mie-type modes, an additional high-quality-factor resonant state appears if some structure asymmetry is created. These states are referred to trapped modes when the electromagnetic field is strongly confined inside the structure being very weakly coupled to free space. We propose to make a proof of concept regarding the trapped mode excitation in a single-particle planar all-dielectric metamaterial considering the sample fabrication with an inexpensive bottom-up-chemical technology where particles are produced from commercially available ceramic materials for operation in the microwave range. The proposed single-particle design of disks with a penetrating notch gives more tuning freedoms where the notch parameters open a prospect to find an optimal solution for the unit cell geometry effectively producing the trapped mode excitation. Moreover, the notch can be filled with other material possessing properties of a nonlinear or gain medium to expand the metamaterial functionality.

\section*{Appendix A: Scattering on a single particle}

\begin{figure}[ht!]
\centering\includegraphics[width=8cm]{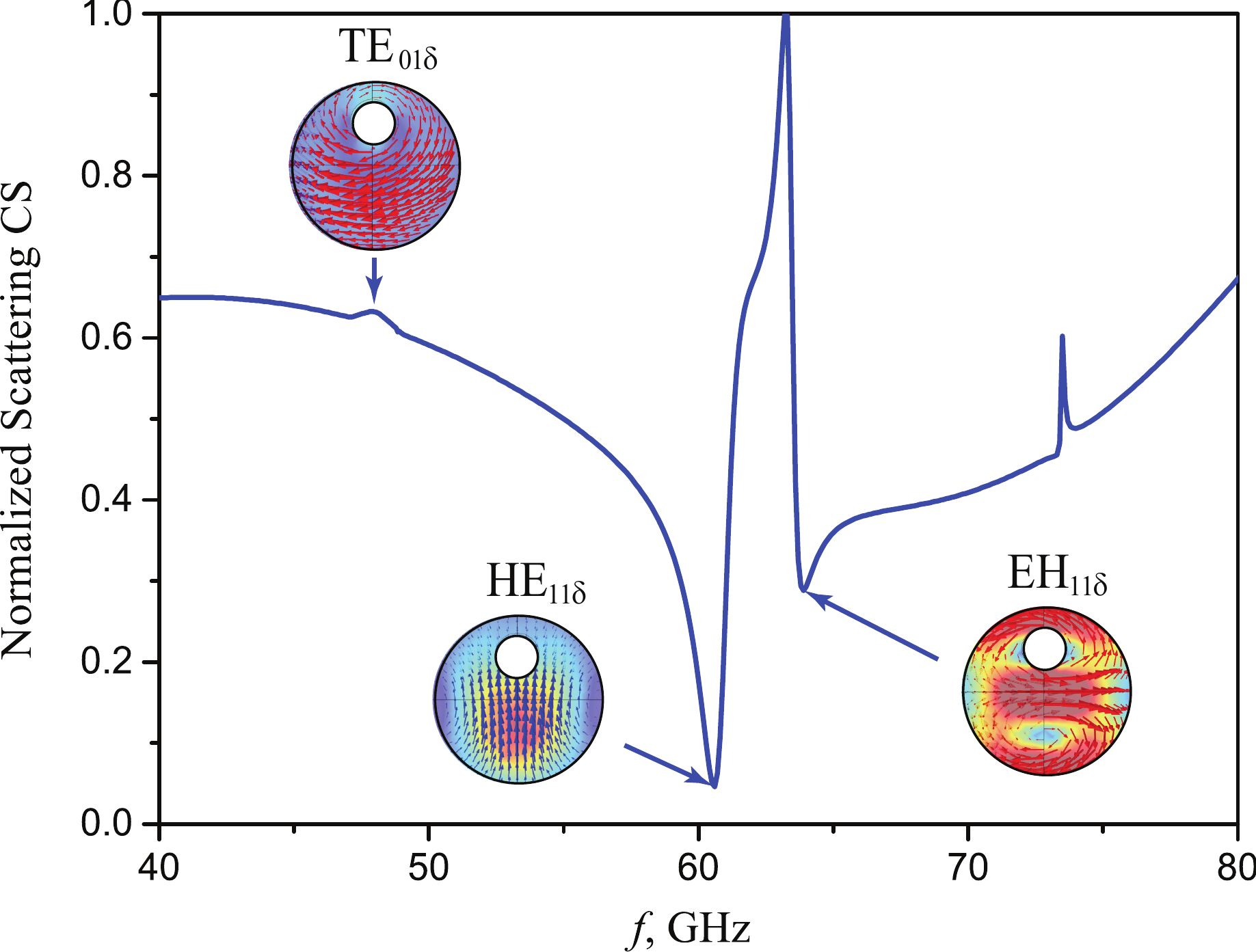}
\caption{Normalized scattering cross section of a dielectric disk having off-centered round penetrating hole made through it; $a_d=0.457$~mm, $h_d=0.417$~mm,  $h_s=0$~mm, $\theta=a_{d}/2$, $a_h=a_{d}/4$.}
\label{fig:scattering}
\end{figure}

We performed additional simulations on a single particle scattering in order to prove that the discussed resonant states are influenced by the modes of an individual dielectric resonator rather than by the electromagnetic coupling (periodic effect) between them. Thus, normalized  scattering cross sections (CS) of a solid disk and a disk possessing off-centered penetrated hole are presented in Fig.~\ref{fig:scattering} for the frequency band of interest. It is supposed that the disk is illuminated by a linearly polarized plane wave propagating along the cylinder's axis (frontal excitation). Insets show patterns of the electric and magnetic fields at the corresponding resonant frequencies, from which one can conclude that they completely correspond to those existed in the unit cell of the metamaterial under study [see Figs.~\ref{fig:symmetric}(b) and \ref{fig:asymmetric}(b)]. For details on the single particle scattering see also Ref.~\cite{Baryshnikova_JOptSocAmB_2017}.

\section*{Acknowledgments}
The authors thank Prof. Yu. S. Kivshar for fruitful discussions and suggestions.

\end{document}